# Giant optical anisotropy in transition metal dichalcogenides for next-generation photonics

*G.A. Ermolaev[1,2], D.V. Grudinin[1], Y.V. Stebunov[3], K.V. Voronin[1,2], V.G. Kravets[4], J. Duan[5,6], A.B. Mazitov[1,7], G.I. Tselikov[1], A. Bylinkin[1,8], D.I. Yakubovsky[1], S.M. Novikov[1], D.G. Baranov[9,1], A.Y. Nikitin[10,11,1], I.A. Kruglov[1,7], T. Shegai[9], P. Alonso-González[5,6], A.N. Grigorenko[4], A.V. Arsenin[1,12], K.S. Novoselov[13,3,14], V.S. Volkov[1,12]\**

[1]Center for Photonics and 2D Materials, Moscow Institute of Physics and Technology, Dolgoprudny 141700, Russia.

[2]Skolkovo Institute of Science and Technology, Moscow 121205, Russia.

[3]National Graphene Institute (NGI), University of Manchester, Manchester M13 9PL, UK.

[4]Department of Physics and Astronomy, University of Manchester, Manchester M13 9PL, UK.

[5]Department of Physics, University of Oviedo, Oviedo 33006, Spain.

[6]Center of Research on Nanomaterials and Nanotechnology, CINN (CSIC-Universidad de Oviedo), El Entrego 33940, Spain.

[7]Dukhov Research Institute of Automatics (VNIIA), Moscow 127055, Russia.

[8]CIC nanoGUNE BRTA, Donostia-San Sebastián 20018, Spain.

[9]Department of Physics, Chalmers University of Technology, Göteborg 412 96, Sweden.

[10]Donostia International Physics Center (DIPC), Donostia-San Sebastián 20018, Spain.

[11]IKERBASQUE, Basque Foundation for Science, Bilbao 48013, Spain.

[12]GrapheneTek, Skolkovo Innovation Center, Moscow 143026, Russia.

[13]Department of Materials Science and Engineering, National University of Singapore, Singapore 117574, Singapore.

[14]Chongqing 2D Materials Institute, Liangjiang New Area, Chongqing 400714, China.

*e-mail: volkov.vs@mipt.ru

**ABSTRACT**

Large optical anisotropy observed in a broad spectral range is of paramount importance for efficient light manipulation in countless devices. Although a giant anisotropy was recently observed in the mid-infrared wavelength range, for visible and near-infrared spectral intervals, the problem remains acute with the highest reported birefringence values of 0.8 in $BaTiS_3$ and h-BN crystals. This inspired an intensive search for giant optical anisotropy among natural and artificial materials. Here, we demonstrate that layered transition metal dichalcogenides (TMDCs) provide an answer to this quest owing to their fundamental differences between intralayer strong covalent bonding and weak interlayer van der Walls interaction. To do this, we carried out a correlative far- and near-field characterization validated by first-principle calculations that reveals an unprecedented birefringence of 1.5 in the infrared and 3 in the visible light for $MoS_2$. Our findings demonstrate that this outstanding anisotropy allows for tackling the diffraction limit enabling an avenue for on-chip next-generation photonics.

**INTRODUCTION**

Optical anisotropy plays a crucial role in light manipulation owing to birefringence phenomena, namely, doubling the incoming light into two different rays (called ordinary and extraordinary for uniaxial optical materials), which results in spatial and polarization separation,[1] through versatile optical components,[2–4] including polarizers, wave plates, multilayer mirrors, and phase-matching elements. Their performance primarily depends on the phase retardance ($\varphi$) between ordinary and extraordinary rays, which is proportional to the thickness ($d$) of the device and the birefringence ($\Delta n$) of the constituting materials. Thus, a large birefringence is highly favorable and beneficial since it leads to more compact and efficient devices. Despite its great importance for high-performance optics, the currently used materials such as inorganic solids and liquid crystals possess a quite small birefringence with typical values below 0.4.[5–9] Even the record-holders quasi-one-dimensional $BaTiS_3$ and layered h-BN crystals improve this result by less than twofold ($\Delta n$ ~ 0.8).[10,11] The problem is partially solved in the mid-infrared range by large anisotropy in the biaxial van der Waals (vdW) crystals α-MoO$_3$ and α-V$_2$O$_5$.[12,13] Still, these materials become mostly isotropic in the visible and near-infrared light. Meanwhile, artificial design can offer large birefringence in metamaterials and metasurfaces.[14] However, its widespread usage is impeded by optical losses and fabrication challenges.

As a result, natural materials with giant anisotropy ($\Delta n > 1$) are in high demand both for scientific and industrial purposes. In this regard, transition-metal dichalcogenides (TMDCs) in a bulk configuration are promising candidates because of their strongly anisotropic vdW structure, which naturally leads to a large intrinsic birefringence. In particular, while MoS$_2$ solids adopt an in-plane crystalline layered structure through strong ionic/covalent bonds between molybdenum (Mo) and sulfur (S) atoms, the out-of-plane link of these layers occurs via weak vdW forces in trigonal prismatic configuration[15], as illustrated in Figure 1a.

As a consequence, a strong optical anisotropy emerges in TMDCs. A diagonal permittivity tensor can describe it with two optical constants corresponding to the crystallographic *ab*-plane and the *c*-axis.[16] Interestingly, these anisotropic properties of TMDCs were qualitatively demonstrated back in 1967 by Liang *et al.*,[17] but only currently attracted significant importance in experiments dealing with novel regimes of light-matter interactions[18,19] comprising exciton-polariton transport,[20] Zenneck surface waves,[21] tunable modal birefringence,[22] and anapole-exciton polaritons.[23] Although a recent pioneering work by Hu and co-workers[16] reported a birefringence value of $\Delta n = 1.4$ for MoS$_2$ at $\lambda = 1530$ nm, the values of asymmetric dielectric responses of MoS$_2$ in a wide wavelength interval have so far remained unknown. Most likely, it stems from inherent experimental difficulties while measuring a high refractive index of anisotropic materials, which we overcome here by joining together far and near-field characterization techniques. The method allows us to obtain the full dielectric tensor in a wide wavelength range (360 – 1700 nm) and reveals giant birefringence for MoS$_2$ as high as $\Delta n \sim 1.5$ in the infrared and $\Delta n \sim 3$ in the visible spectra. This outstandingly large optical anisotropy accompanied by a high refractive index $n \sim 4$ paves the way for highly efficient optics and light manipulation in photonic chips.

**RESULTS**

The measurement of the anisotropic optical response of TMDCs is a challenging task because of multiple experimental obstacles. One of the hardest parts is the implementation of traditional spectroscopic diffraction-limited characterization techniques, including transmittance, reflectance, and ellipsometry measurements, since bulk TMDCs are usually prepared by the exfoliation method, as a result, the samples obtained have lateral dimensions of tens micrometers. A second difficulty is related to the measured signal's low-sensitivity to the out-of-plane components attributed to a large in-plane refractive index $n \sim 4$. For instance, in ellipsometry configuration, incident light at 80° gives the refraction angle of only 14.3° according to the Snell's law, which implies that the probed electric field of the refracted light mainly lies along the layers and thus, is almost insensitive to the out-of-plane dielectric component.

To overcome the latter, we prepared an exfoliated thin $MoS_2$ films on 285 nm $SiO_2$/Si substrate and verified their 2H semiconducting configuration by the resonant Raman demonstrated in the inset of Figure 1a since $MoS_2$ exists in nature in three phase modifications (see details in Supporting Information): semiconducting (2H and 3R) and metallic (1T).[24] The thick layer of silicon oxide will produce an interference-like pattern for ordinary and extraordinary beams, which can readily be detected employing phase-sensitive techniques such as spectroscopic ellipsometry (SE). For this reason, we performed imaging spectroscopic ellipsometry (ISE) measurements in the 360 – 1700 nm wavelength range, given that it allows measuring samples down to several micrometers since it is a hybrid of ellipsometer and microscope (Methods). It allowed us to record the ellipsometry signal Ψ and Δ (Methods) from several regions of interest (ROI) of the flakes within the selected field of view, as indicated in Figure 1b. As a result, multiple sample analysis was implemented to increase data reliability (see Supporting Information). The resulting ellipsometry spectra in Figure 1c indeed shows a pronounced asymmetrical interference-like peak at around 900 nm. This is induced by a large phase difference between ordinary and extraordinary beams, indicating (without any modeling of the experimental curves) a large birefringence stemming from a strong anisotropy between the *c*-axis and the *ab*-plane.

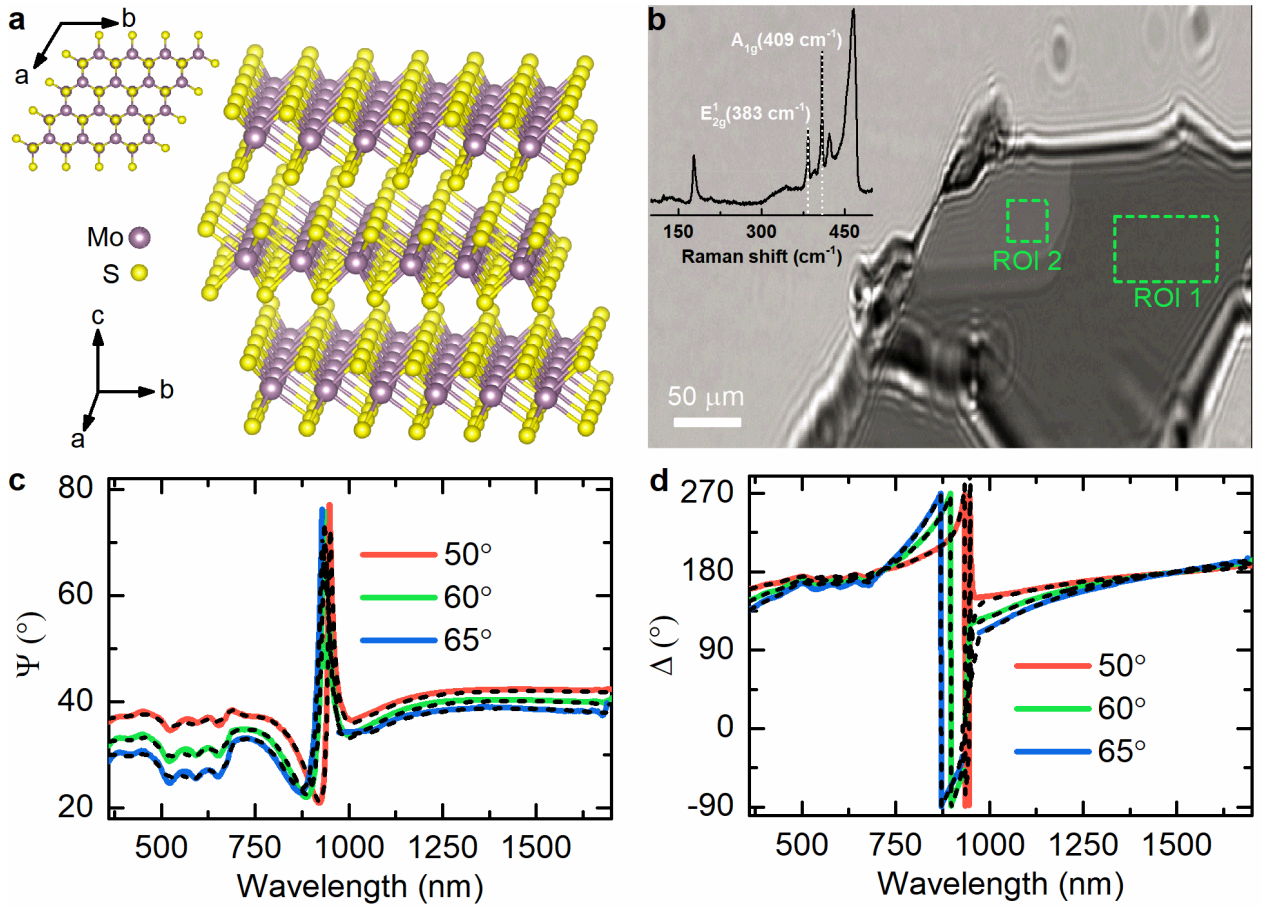

**Figure 1. Anisotropy in MoS₂. a** Schematic illustration of the $MoS_2$ layered structure: the giant anisotropy between *ab*-plane and *c*-axis arises from different interlayer (weak van der Waals bonding) and intralayer (strong ionic/covalent bonding) atomic interactions. **b** Optical ellipsometer microscope image of the exfoliated $MoS_2$ thin film on a 285 nm-thick $SiO_2$/Si substrate at 65°. The ellipsometry measurements were performed in the two uniform areas marked by the green dashed lines. The inset shows a resonant Raman spectrum at an excitation wavelength $\lambda = 632.8$ nm with the characteristic modes $E_{2g}^1 = 383$ cm⁻¹ and $A_{1g} = 409$ cm⁻¹, whose positions confirm the 2H semiconducting material configuration. **c-d** Experimental (solid lines) and analytically calculated (dashed lines, see Methods) ellipsometric parameters Ψ and Δ for ROI 1 (Ψ and Δ for ROI 2 see in Supporting Information) at three incident angles 50°, 60°, and 65°. The asymmetric interference-like peak at around 900 nm is induced by interference enhancement in $SiO_2$ caused by splitting the incident beam into ordinary and extraordinary beams indicating a giant anisotropy in $MoS_2$.

Notwithstanding the noticeable anisotropic feature at 900 nm in the measured spectra, to accurately retrieve the complete dielectric tensor of $MoS_2$ and enable predictive capabilities for future advanced optical devices using this material, it is imperative to develop an accurate dielectric function model. In that case, the best route towards a dielectric description is to utilize the crystallographic features of $MoS_2$. Briefly, in its 2H structure consecutive layers are linked by weak vdW forces and rotated by 180° with respect to each other leading to a strong suppression of interlayer hopping for both electrons and holes, and, thus, preventing the formation of tightly bound interlayer electron-hole pair upon light illumination, the so-called excitons.[25,26] Therefore, along the *c*-axis, the material is transparent, and a Cauchy model describes its dielectric response (see Supporting Information), which is an evident consequence of Kramers-Kronig relation between real (*n*) and imaginary (*k*) parts of the refractive index and material transparency. In contrast, the confinement of electron and holes within the layer results in enormous binding energy (~ 50 meV) for intralayer A- and B- excitons at the visible range similar to its monolayer counterpart.[27] At the same time, it supports C and C' exciton complexes at ultraviolet wavelengths due to the nest banding effects and complex atomic orbital contributions.[28] The Tauc-Lorentz oscillator model best describes this excitonic behavior for *ab*-plane (see Supporting Information) because it captures two the most essential physical features:[29] (*i*) at low photon energies, excitons cannot be excited, as a consequence, absorption, or equivalently the imaginary part of refractive index (*k*), is equal to zero in this wavelength range and (*ii*) excitonic peaks exhibit an asymmetric shape due to phonon coupling of bright (excited by light) and dark (not excited by light) excitons.

Using these models for describing the optical properties of $MoS_2$, we fitted the experimentally measured ellipsometric parameters Ψ and Δ for both ROIs at the same time (see Supporting

Information). The resulting ordinary (along the layers) and extraordinary (perpendicular to the layers) optical constants and birefringence are displayed in Figure 2a and have a surprisingly well match with the values predicted by the first-principle calculations (see Methods and Supporting Information). As expected, the material along the $c$-axis is transparent, even at ultraviolet and visible wavelengths. It confirms that excitons are formed in the layers and provide a dichroic window from ~ 900 nm where the absorption of both ordinary and extraordinary light becomes negligible. Of immediate interest is the giant birefringence of $\Delta n$ ~ 1.5 in the infrared and $\Delta n$ ~ 3 in the visible ranges, which can serve as a platform for optical engineering in creating of novel devices for the photonic application. As compared in Figure 2c, the birefringence obtained for $MoS_2$ in the visible and near-infrared spectral intervals is several times larger than for previous record-holders $BaTiS_3$ and h-BN,[10,11] and an order of magnitude exceeding the values of currently used birefringent materials. Particular attention should also be given to the absolute values of the refractive indices, specifically their in-plane component. The high value of ~ 4.1 is comparable with traditionally used isotropic high-refractive-index semiconductors,[30] including Si (~ 3.6),[31] GaAs (~ 4.3),[32] and GaSb (~ 3.9)[33] as illustrated in Figure 2b. Such a large refractive index for $MoS_2$ opens the door for lossless subwavelength photonics with the resolution of ~ 100 nm, which can easily rival with plasmonics platform, yet does not suffer from losses.

Furthermore, as we have not specified particular properties of $MoS_2$ other than its in-plane excitonic nature and the out-of-plane transparency, our conclusions are quite general, applying equally well to other semiconductor members of layered TMDCs with hexagonal, tetragonal, and trigonal atomic structure.[1] Consequently, we anticipate that other TMDCs with hexagonal configuration also exhibit giant anisotropy because of the crystal's similarity. In fact, the refractive index depends upon the density of atoms in the crystal. As a consequence, naively, one could expect that the larger the distance between the layers, the higher the birefringence value between in-plane and out-of-plane dielectric response. Indeed, a comparison of their lattice parameters normalized to the distance between chalcogen intercore distances in the inset of Figure 2b explains the giant anisotropy in MoS$_2$ and forecasts the similar (or even higher) birefringence values for MoSe$_2$, WS$_2$, WSe$_2$, and WTe$_2$.

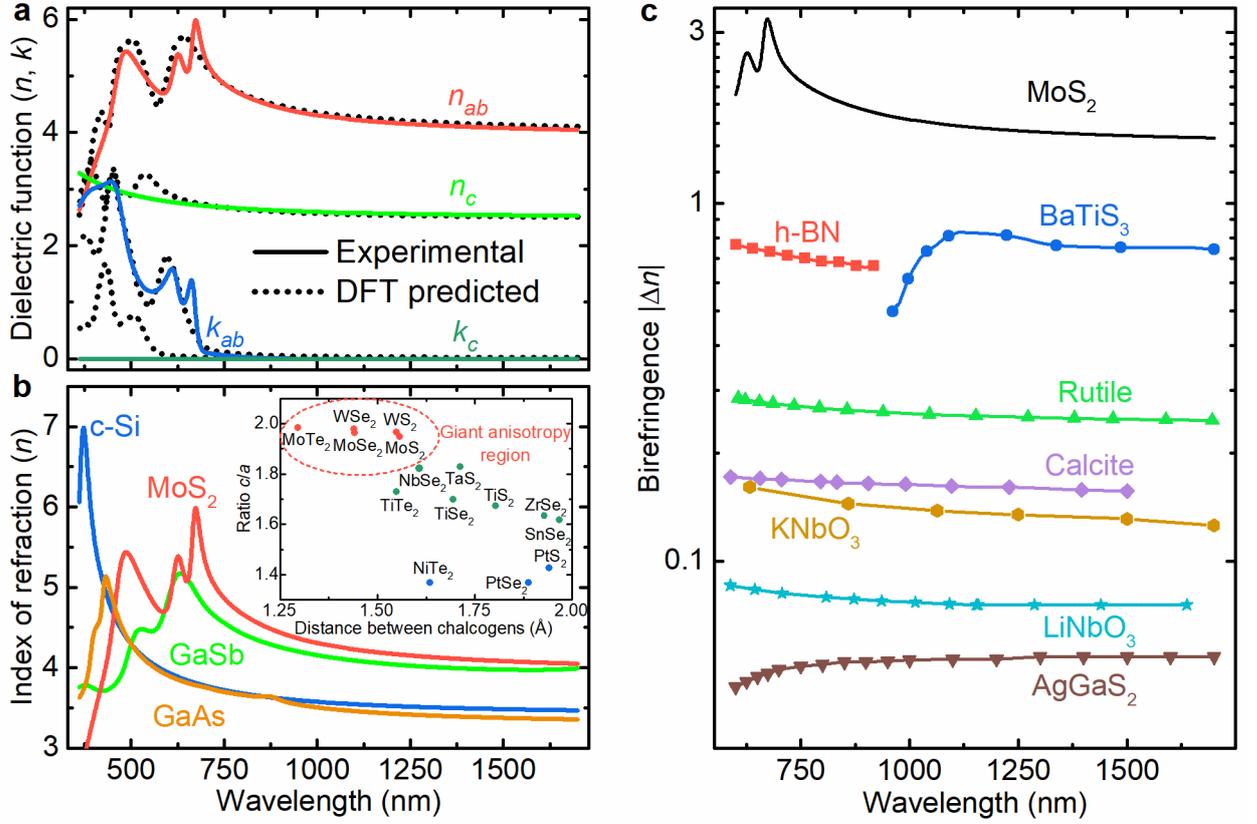

**Figure 2. Optical anisotropy of MoS$_2$. a** Real ($n$) and imaginary ($k$) parts of the dielectric function along the *ab*-plane and *c*-axis. **b** Comparison of the MoS$_2$ refractive index along the *ab*-plane with other high-refractive-index materials commonly used in nanophotonics.[31–33] The inset compares the ratio of crystal parameters ($c$ and $a$ in Figure 1a) versus distance between neighboring chalcogens (S, Se, and Te) for various TMDCs. The highest anisotropy is expected for TMDCs denoted by red circles, while the lowest for blue. The crystallographic data were adopted from the review article.[34] **c** Comparison of the absolute birefringence values of MoS$_2$ with different birefringent materials, including h-BN and BaTiS$_3$, reported showing the highest anisotropy so far. The birefringence values for other materials in (**c**) were adopted from several reports.[5–11]

For an unambiguous validation of the extracted dielectric function, we analyzed the planar transverse magnetic (TM) waveguide modes propagating in MoS$_2$ flakes employing a scattering-type scanning near-field optical microscope (s-SNOM, Methods). By recording the scattered radiation, nanoscale images corresponding to the field distribution associated with the guided mode is obtained (Figure 3a-b). The effective TM-waveguide mode index ($n_{\text{eff,TM}}$) strongly depends on the material anisotropy allowing to probe anisotropic response, in-plane ($n_{ab}$) and out-of-plane ($n_c$) refractive indices, and determined by:[16]

$$\frac{2\pi d}{\lambda}\sqrt{n_{ab}^2 - n_{\text{eff,TM}}^2 \frac{n_{ab}^2}{n_c^2}} = \tan^{-1}\left(\frac{n_{ab}^2}{n_{\text{Air}}^2}\frac{\sqrt{n_{\text{eff,TM}}^2 - n_{\text{Air}}^2}}{\sqrt{n_{ab}^2 - n_{\text{eff,TM}}^2 \frac{n_{ab}^2}{n_c^2}}}\right) + \tan^{-1}\left(\frac{n_{ab}^2}{n_{\text{SiO}_2}^2}\frac{\sqrt{n_{\text{eff,TM}}^2 - n_{\text{SiO}_2}^2}}{\sqrt{n_{ab}^2 - n_{\text{eff,TM}}^2 \frac{n_{ab}^2}{n_c^2}}}\right) + m\pi, \qquad (1)$$

where $d$ is the thickness of the MoS$_2$ flake, $\lambda$ is the incident wavelength, $n_{\text{Air}} = 1$ and $n_{\text{SiO2}} = 1.45$ are air and SiO$_2$ refractive indices, and $m$ is the mode order. We used incident wavelengths in the range 1470 – 1570 nm and 632.8 nm to excite guiding modes by focusing light into the apex of the s-SNOM tip, which allows for momentum matching conditions. The excited mode propagates in the MoS$_2$ nanoflake as cylindrical waves, which interfere with the illuminating plane wave giving rise to interferometric patterns of the near-field,[20] as clearly seen in Figure 3c-d. It is worth mention that while most of the previous works with s-SNOM for TMDCs focus only on the near-field amplitude,[16,20] the most accurate results are obtained by analyzing the phase as well[35] (see Supporting Information for comparison). To retrieve the effective waveguide mode index, in Figure 3e, we analyzed the Fourier transform (FT) of individual line scans from Figure 3d. The resulting FT has two pronounced peaks: one around zero due to background originating mostly from a strong tip-sample coupling, and the second one associated with the planar TM-waveguide mode of interest. Note that there are no peaks in the left part of Figure 3e (for negative values of $q$), indicating that no modes propagate in the backward direction (from the edge to the tip). The latter implies that mode scattering by the edge is far more efficient than mode edge reflection or launching. Otherwise, we would observe standing waves with a cosine form, whose FT would be symmetrical and which are predominantly observed in nano-infrared imaging of graphene plasmons[36,37] and hexagonal boron nitride (hBN) polaritons.[38] The primary reason for the observed tip-launching and edge-scattering mechanisms is the relatively small momenta of the modes[20] since it is much closer to the free-space photon wavevector ($k_0$) than in the studies of graphene[36,37] or hBN.[38] For small momenta, the effective mode index is connected with that determined from the FT ($n_{\text{s-SNOM,FT}}$) by momentum conservation along the edge direction:[20]

$$n_{\text{eff,TM}} = n_{\text{s-SNOM,FT}} + \cos(\alpha) \cdot \sin(\beta) \quad (2)$$

where in our case $\alpha = 45°$ is the angle between the illumination wavevector and its projection $k_\parallel$ on the sample surface plane and $\beta = 80°$ is the angle between $k_\parallel$ and the sample edge. Based on the extracted $n_{\text{eff,TM}}$, we constructed the energy ($E = h \cdot c/\lambda$)–momentum ($q = 1/\lambda$) dispersion relation of the waveguide mode. The obtained experimental ($q, E$) data points (green triangles) are overlaid on top of the calculated dispersion color map in Figure 4a using constants from Figure 2a. For reference, we also added the dispersion for an isotropic model in Figure 4b, assuming the optical constants to be the same for all crystallographic axes and equal one from *ab*-plane. Notably, in the visible spectral range, where excitons start playing a role, the isotropic model (Figure 4b) predicts the absence of guided modes owing to high material absorption. Conversely, our anisotropic results and near-field measurements reveal that even for this spectral interval, guided modes exist, which explains the recently discovered excitons polaritons in TMDCs.[20] Therefore, the excellent agreement between the experiment and theory validates our dielectric permittivity of $MoS_2$, allowing for predicting capabilities in future photonic devices, including polarization-maintaining fibers[4] and polarization tunable Mie-nanoresonators for nonlinear photonics[39] since the magnetic dipole (MD) Mie-resonance in $MoS_2$ nanoparticles is strongly affected by the refractive index values.[40] For instance, the spectral position of MD-resonance for a spherical particle is approximately defined by $\lambda_{\text{MD}} \approx nD$, with $D$ being the sphere's diameter.[41] Besides, its anisotropic behavior allows its use as nanoresonators even for photon energies higher than the electronic bandgap thanks to the absence of absorption along the *c*-axis, while for conventional isotropic materials (c-Si, GaAs, and GaSb) this option is closed. Therefore, TMDCs provide device miniaturization, and their birefringence enables fine-tuning the resonance position in a wide spectral range by altering the light polarization, which is roughly $n_{ab}D - n_cD \approx 450$ nm for a typical diameter of 300 nm.

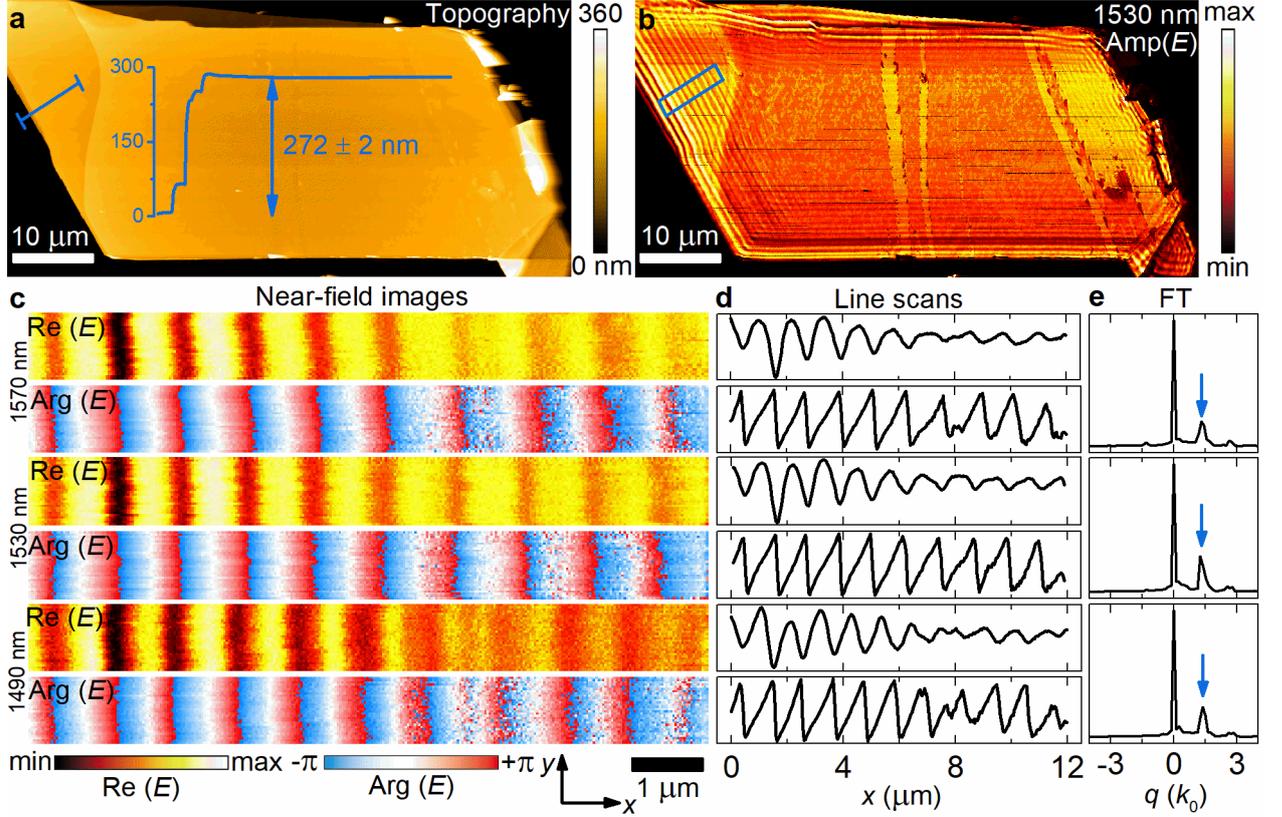

**Figure 3. Waveguide modes imaged by s-SNOM. a** Topography image of the analyzed $MoS_2$ flake. A profile taken from the region labeled by a blue line is shown. **b** Near-field amplitude image of the flake recorded at $\lambda = 1530$ nm. The region with the strongest signal is framed by a blue rectangular. **c** Near-field images, real part Re($E$) and phase Arg($E$), of the electric field $E$ taken at 1570 nm (top), 1530 nm (middle), and 1490 nm (bottom) in an area of the image in **(b)**, indicated by a blue rectangle. **d** $x$-line scans taken from **(c)** and averaged over 1.2 μm along the $y$-axis (other wavelength images are collected in Supporting Information). **e** Fourier transform (FT) amplitude of the complex near field signal in **(d)**, the blue arrow marks the peak associated with the waveguide mode.

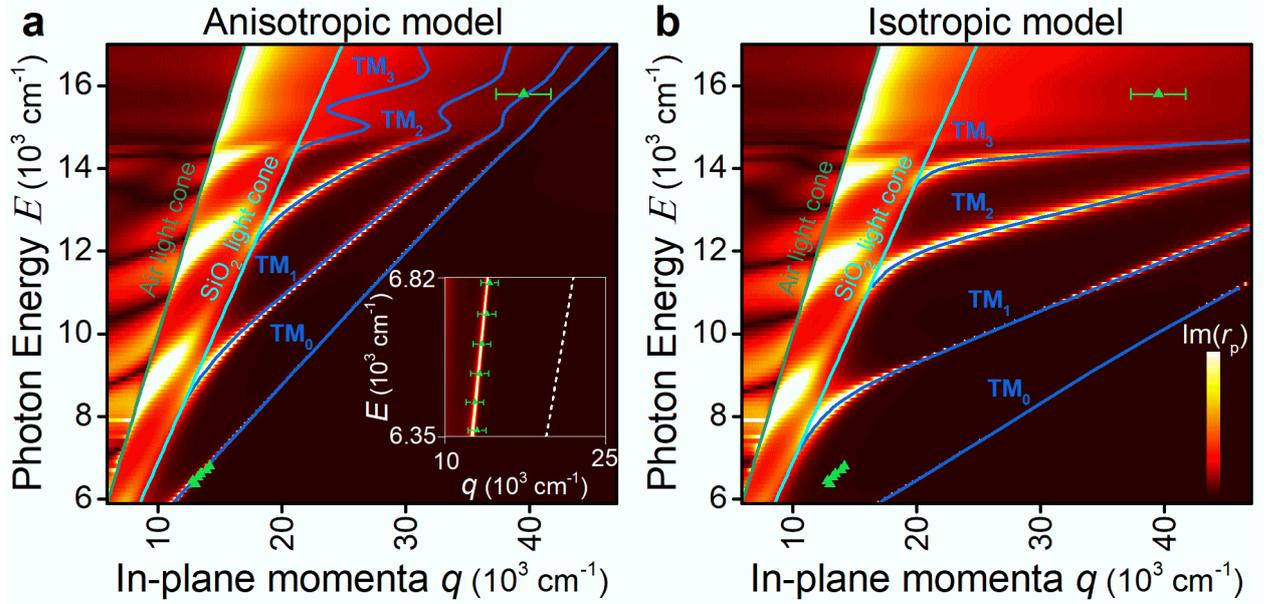

**Figure 4. Dispersion of a planar MoS$_2$ waveguide. a-b** Transfer matrix calculations[42] for the MoS$_2$/SiO$_2$/Si system for anisotropic and isotropic (in the assumption of optical constants for the *c*-axis equal to those for *ab*-plane) models of MoS$_2$. The experimental ($q = 1/\lambda$, $E = h \cdot c/\lambda$) data points (green triangles) show good agreement with the calculated dispersion (blue lines) based on anisotropic dielectric function from Figure 2a. The inset is a magnified near-infrared spectral range with a white dashed line illustrating the isotropic case's dispersion relation.

More importantly, the giant anisotropy provides an avenue to resolve the fundamental limitations of isotropic materials. In this regard, extreme skin depth (e-skid) waveguide in Figure 5 epitomizes the importance of highly anisotropic material as an essential building block for next-generation photonics.[43] This phenomenon originates from the general law of total internal reflection (TIR). In the classical case in Figure 5a, if $n_1 > n_2$ ($n_1$ and $n_2$ are the refractive indices of the medium 1 and 2) and the incident angle is greater than the critical angle, light is reflected in the first medium and decays in the second. This effect is in the core of telecommunication technologies for guiding light at vast distances since light reflects without power loss. However, reducing the thickness of standard (typically Si) optical waveguides down to the nanometric scale introduces a big difficulty: the confinement of the electromagnetic modes in the optically less dense material significantly weakens, as illustrated in Figure 5a-b and e-f. It restricts the miniaturization on photonic integrated circuits because of the cross-talk (electric field overlapping between adjacent waveguides).[44] To date, the best solution to the problem is the use of a generalized form of TIR at the interface of isotropic and uniaxial metamaterial asserted that $n_1$ should be larger than out-of-plane component $n_c$ (Figure 5c).[45] Counterintuitively, the in-plane component $n_{ab}$ could have any value even higher than $n_1$ and, more interestingly, the greater its magnitude, the better the light confinement inside the waveguide core (Figure 5d), allowing to get closer to the diffraction limit ($\lambda/2n_{core}$). To experimentally demonstrate the effect, we covered 285 nm-thick MoS$_2$ flake with 190 nm-thick silicon (Methods) to form a planar e-skid waveguide Air/Si/MoS$_2$. Although MoS$_2$ is better than air for light confinement, we left one of the faces of the silicon slab uncovered to visualize the mode by near-field measurements shown in Figure 5i-l. The waveguiding mode measured dispersion is in close agreement with the theoretical calculations, thus validating the e-skid waveguide concept for light confinement and, consequently, miniaturized photonic integrated circuits. It is worth mention that MoS$_2$ yields markedly better light confinement than recently introduced metamaterial[43] of alternating layers of Si and SiO$_2$ (Figure 5f) since MoS$_2$ is a natural metamaterial with molybdenum, sulfur and vdW gap varying layers (Figure 5c-d). Finally, it leads to another concept of a vertical integrated circuit, which recently has been proved to be a useful degree of freedom for efficient light manipulation.[46]

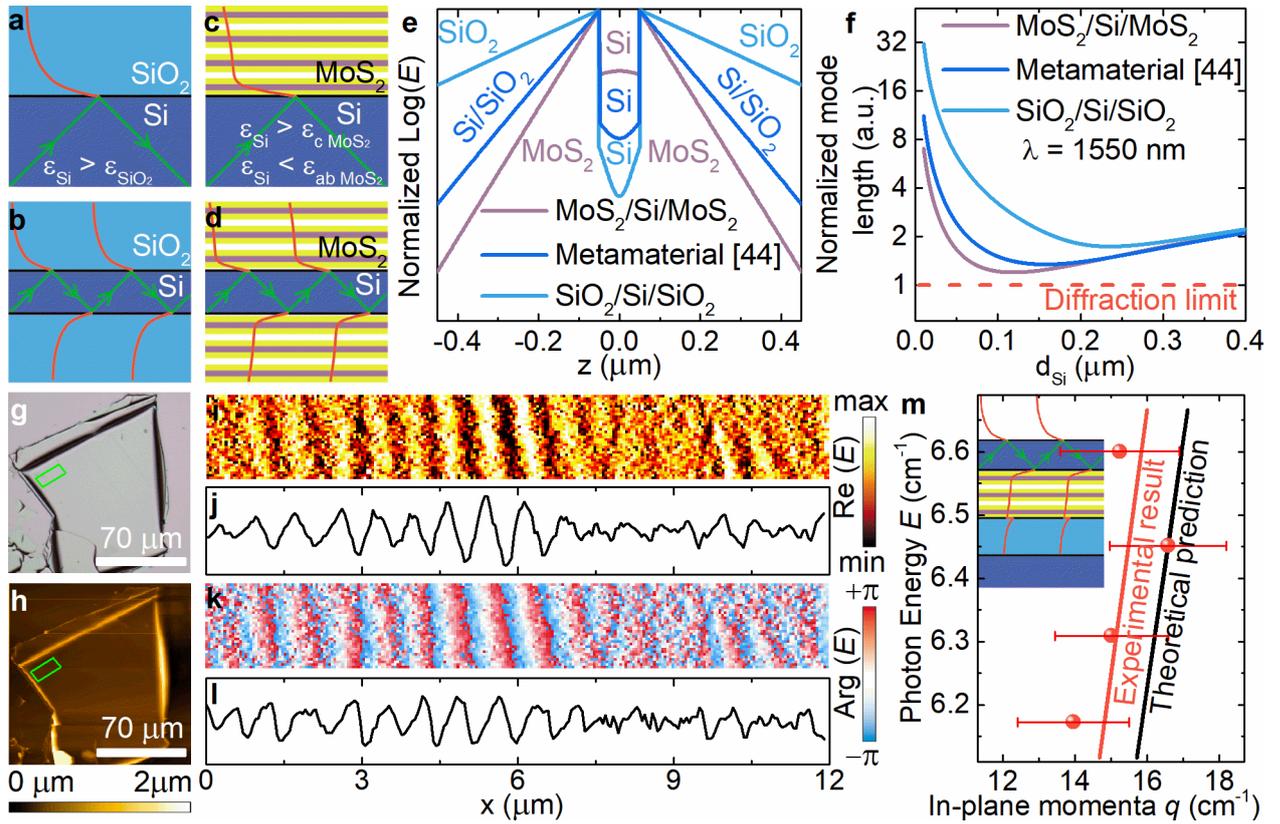

**Figure 5. Transparent sub-diffraction optics. a** Traditional total internal reflection with two isotropic media: above the critical angle, the light is reflected from the interface giving a decaying wave for a lower refractive index medium. **b** It results in long electric tails in conventional waveguides limited the current on-chip photonics. **c** Relaxed total internal reflection: the out-of-plane dielectric component is less than for isotropic material, while the high in-plane part provides substantial light compression. **d** It gives rise to a planar waveguide with outstandingly short electric tails. Red curves for **(a-d)** panels schematically shows electric field amplitude, and yellow, purple, and white colors in **(c-d)** panels label sulfur, molybdenum, and vdW gap layers. **e** Comparison of electric field distribution for $MoS_2/Si/MoS_2$, $SiO_2/Si/SiO_2$ metamaterial/Si/metamaterial[45] systems reveals that giant anisotropy causes giant light compression. **f** Light confinement in $MoS_2/Si/MoS_2$ allows for tackling the diffraction limit compared with traditionally used $SiO_2/Si/SiO_2$ and recently introduced[45] metamaterial cladding with alternating layers of Si and $SiO_2$ using silicon as a core. **g-h** Optical image and AFM topography mapping of the flake with 190 nm covered silicon. **i-l** Near field images, real part Re ($E$) and phase Arg ($E$), with the corresponding line scans of the electric field taken at $\lambda = 1550$ nm from the area of **(g-h)** indicated by a green rectangular. (other

wavelength images are collected in Supporting Information) **m** Comparison between theoretical and experimental dispersion. The inset is an artistic representation of the investigated system Air/Si (195 nm)/MoS$_2$ (285 nm)/SiO$_2$ (285 nm)/Si.

## DISCUSSION

Optical anisotropy lies behind the functionality of many optical devices such as polarizers and wave plates, to name a few. This phenomenon's outstanding importance leads to an active investigation of anisotropic materials and the expansion of their application scope. However, the apparatus efficiency and compactness mostly depend on absolute birefringence values, which are moderate with the best result ($\Delta n < 0.8$), reported in h-BN and BaTiS$_3$ crystals in the visible and near-infrared ranges. We believe that these limitations can be outperformed by the family of TMDCs materials, whose inherent intralayer excitonic behavior results in large anisotropy along and perpendicular to the layers. To validate the concept, we have shown a giant ($\Delta n > 1.5$) broadband anisotropy for MoS$_2$ employing far and near-field techniques. Additionally, we demonstrated its applicability for on-chip sub-duffraction optics. From a wider perspective, our result establishes new avenues for next-generation nanophotonics based on TMDCs, for example, in tunable Mie-nanoresonators, and exciton-polariton physics.

## ACKNOWLEDGEMENTS

The authors thank Dr. Sebastian Funke, Dr. Shun Okano, Dr. Matthias Duwe, Natalia Doroshina, and Valentyn Solovey for their help in performing ellipsometry and Raman spectra. The authors also thank the MIPT Language Training and Testing Center (LTTC), particularly, Dr. Elena Bazanova, for the help with English language editing. We gratefully acknowledge financial support from the Ministry of Science and Higher Education of the Russian Federation (No. 0714-2020-0002) and the Russian Foundation for Basic Research (20-07-00475, 18-29-02089 and 20-07-00840). A.Y.N. acknowledges the Spanish Ministry of Science, Innovation and Universities (national project no. MAT201788358-C3-3-R) and the Basque Department of Education (grant

PIBA-2020-1-0014). P.A-G. and J.D. acknowledge financial support from the European Research Council under Starting Grant 715496, 2DNANOPTICA. A.N.G., V.G.K. and K.S.N. acknowledges support from EU Graphene Flagship Core 3 (881603). K.S.N. also acknowledges support from EU Flagship Program 2D-SIPC Quantum Technology, European Research Council Synergy Grant Hetero2D, the Royal Society, EPSRC grants EP/N010345/1, EP/P026850/1, EP/S030719/1.

**METHODS**

**Sample preparation.** The $MoS_2$ microcrystals were exfoliated on silicon wafers with 285-nm-thick thermal $SiO_2$ from a synthetically grown bulk $MoS_2$ sample purchased from the 2D Semiconductors Inc.

**Ellipsometry setup.** Imaging spectroscopic ellipsometry (ISE) measurements were performed with a commercial spectroscopic nulling ellipsometer EP4 (https://accurion.com). Spectroscopic data are obtained in the spectral range 360 – 1700 nm in step with 1 nm. The light is guided through for linear polarization and then through a compensator to prepare elliptically polarized collimated light so that the reflected light from the sample is again linearly polarized. The reflected light is directed through a 10x objective to a CCD camera (microscope configuration). In a suitable coordinate system, the complex reflectance matrix is described by $\tan(\Psi)\cdot\exp(i\Delta)$. The analytical $\Psi$ and $\Delta$ are calculated using Fresnel formulas.[1]

**Near-field optical nano-spectroscopy.** The nano-imaging recording was performed using a commercial s-SNOM (www.neaspec.com). The s-SNOM is based on a tapping-mode AFM illuminated by a monochromatic tunable laser of the wavelength from 1470 – 1570 nm spectral interval or He-Ne laser with the wavelength 632.8 nm. The near-field images were registered by pseudo-heterodyne interferometric module with tip-tapping frequency around 270 kHz with an amplitude of about 40 nm. The noise was significantly suppressed by demodulating the optical signal with a pseudo-heterodyne interferometer at high harmonics, $n\Omega$ (in our case third harmonics).

**Raman spectroscopy.** The experimental setup used for Raman measurements was a confocal scanning Raman microscope Horiba LabRAM HR Evolution (https://www.horiba.com/). The measurements were carried out using linearly polarized excitation at a wavelengths of 532 and 632.8 nm, 300 lines/mm diffraction grating, and ×100 objective (N.A. = 0.90), whereas we used unpolarized detection to have a significant signal-to-noise ratio. The spot size was ~ 0.43 μm. The Raman spectra were recorded with 0.26 mW incident powers and an integration time of 10 s.

**First-principle calculations.** Optical properties of 2H-$MoS_2$ were calculated using density functional theory (DFT) within the generalized gradient approximation[46] (Perdew-Burke-Ernzerhof functional) and the projector-augmented wave (PAW) method[47] as implemented in the Vienna Ab initio Simulation Package (VASP).[48] A two-step approach was used: first, $MoS_2$ crystal structure was relaxed, and a one-electron basis set was obtained from a standard DFT calculation; second, micro- and macroscopic dielectric tensors were calculated using GW approximation. Plane wave kinetic energy cutoff was set to 400 eV, and the Γ-centered k-points mesh sampled the Brillouin zone with a resolution of $2\pi\cdot0.05$ Å$^{-1}$.